\documentclass[]{aa}  
\usepackage{graphicx}
\usepackage{txfonts}
\usepackage{natbib}

\def\figwidth{0.5\linewidth}
\begin{document}
   \title{Thermal phase curves of nontransiting terrestrial exoplanets} 
   \subtitle{1. Characterizing atmospheres}

   \author{F. Selsis
          \inst{1,2} 
          \and
         R. D. Wordsworth
         \inst{3}
         \and
         F. Forget
         \inst{3}
          }

  \authorrunning{F. Selsis et al.}
 \titlerunning{Thermal phase curves of non-transiting terrestrial exoplanets. 2. Atmospheres}

  \institute{
    Universit\'e de Bordeaux, Observatoire Aquitain des Sciences de  
l'Univers, BP 89, F-33271 Floirac Cedex, France \\
 \email{selsis@obs.u-bordeaux1.fr} \and
    CNRS, UMR 5804, Laboratoire d'Astrophysique de Bordeaux, BP 89, F-33271 Floirac Cedex, France \and
Laboratoire de M\'et\'eorologie Dynamique, Institut Pierre Simon Laplace, Paris, France \\
             \email{Robin.Wordsworth@lmd.jussieu.fr, francois.forget@lmd.jussieu.fr}}

  \date{Received February 7, 2011; accepted April 25, 2011}

 
  \abstract
   {Although transit spectroscopy is a very powerful method for studying the composition, thermal properties, and dynamics of exoplanet atmospheres, only a few transiting terrestrial exoplanets will be close enough to allow significant transit spectroscopy with the current and forthcoming generations of instruments. Thermal phase curves (variations in the apparent  infrared emission of the planet with its orbital phase) have been observed for hot Jupiters in both transiting and nontransiting configurations, and have been used to put constraints on the temperature distribution and atmospheric circulation. This method could be applied to hot terrestrial exoplanets.}
   {We study the wavelength and phase changes of the thermal emission of a tidally-locked terrestrial planet as atmospheric pressure increases. We address the observability of these multiband phase curves and the ability to use them to detect atmospheric constituents.}
   {We used a 3D climate model (GCM) to simulate the CO$_2$ atmosphere of a terrestrial planet on an 8-day orbit around an M3 dwarf and its apparent infrared emission as a function of its orbital phase. We estimated the signal to photon-noise ratio in narrow bands between 2.5 and 20~$\mu$m for a 10~pc target observed with a 6~m and a 1.5~m telescope (respectively the sizes of JWST and EChO).}
   {Atmospheric absorption bands produce associated signatures in what we call the variation spectrum. Atmospheric windows probing the near surface atmospheric layers are needed to produce large, observable phase-curve amplitudes. The number and transparency of these windows, hence the observability of the phase curves and the molecular signatures, decreases with increasing pressure. Planets with no atmosphere produce large variations and can be easily distinguished from dense absorbing atmospheres.  }
   {Photon-noise limited spectro-photometry of nearby systems could allow us to detect and characterize the atmosphere of nontransiting terrestrial planets known from radial velocity surveys. Two obvious impediments to these types of observations are the required photometric sensitivity ($10^{-5}$) over the duration of at least one orbit (8-days in the studied case) and the intrinsic stellar variability. However, overcoming these obstacles would give access to one order of magnitude more targets than does transit spectroscopy. }

\titlerunning{Spectral phase curves of nontransiting terrestrial exoplanets}
   \maketitle
%

\section{Introduction}
The atmospheric properties of transiting planets can be studied through eclipse spectroscopy. The molecular composition can be constrained by "transmission" spectroscopy during the primary transit or by emission spectroscopy at the secondary eclipse. This has been achieved from both space and the ground for hot Jupiters \citep[e.g.][transmission, Spitzer]{2007Natur.448..169T}, hot Neptunes \citep[e.g.][emission, Spitzer]{2010Natur.464.1161S}, and more recently for GJ1214b, a 7 M$_{\oplus}$ planet \citep[e.g.][transmission, VLT]{2010Natur.468..669B}. Transits also provide constraints on the thermal structure of the atmosphere \citep[e.g][emission, Spitzer]{2008ApJ...673..526K}, as well as on circulation \citep[e.g.][transmission,VLT]{2010Natur.465.1049S}. The thermal phase curves of transiting planets (variation in the apparent  infrared emission of the planet with the orbital phase) have been observed in a single band \citep[e.g.][Spitzer]{2009ApJ...703..769K} or at different wavelengths \citep[e.g.][Spitzer]{2009ApJ...690..822K}, providing constraints on the day/night brightness temperatures and dynamics.

 \begin{figure*}
\centering
\includegraphics[width=\figwidth]{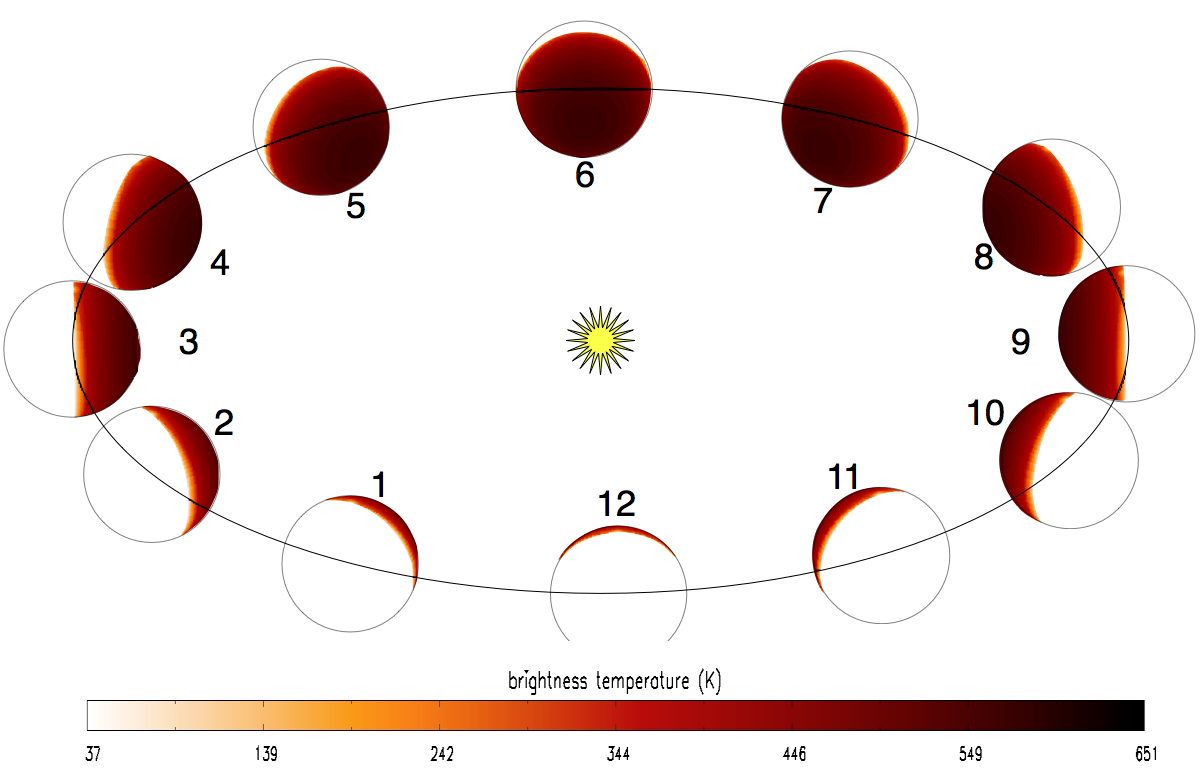}
 \caption{The orbit and phases as seen by the observer. The inclination is set to $60^{\circ}$. The phases shown here are numbered from 1 to 12 and correspond to the phases presented in Fig.\ref{fig:mapflux} to \ref{fig:tbright}.}
      \label{fig:orbit}%
    \end{figure*}
However, applying these methods to characterize (or simply detect) the atmosphere of terrestrial exoplanets with the forthcoming generation of telescopes requires the discovery of transiting systems very close to the Sun (roughly within 10~pc). Short-period planets ($<50$ days) in the $5-20$ Earth mass range have been found by the HARPS radial velocity survey around $30 \pm 10$\% of the G and K stars \citep{2009A&A...493..639M}. For the time being, the closest transiting planet with a mass lower than 20~M$_{\oplus}$ is GJ1214b, found by the MEarth survey at a distance of 13~pc \citep{2009Natur.462..891C}. Statistically, 
there should be one transiting planet from this population for distances between 8 and 18 pc (depending upon orbital periods), and possibly as close as 5~pc if we assume a similar population around M dwarfs. Less than a handful of terrestrial targets will thus be suitable for atmosphere characterization through transit spectroscopy with JWST \citep{2011A&A...525A..83B}. Within this population of short-period, low-mass planets, there are $\sim$10 times more nontransiting planets then transiting ones. A technique able to characterize nontransiting planets and their atmospheres would thus provide enough targets to explore the diversity of terrestrial exoplanet atmospheres.\\
Ambitious projects aimed at directly detecting of terrestrial planets, such as Darwin \citep{2009ExA....23..435C}, TPF-I \citep{2007STIN...0814326L}, TPF-C \citep{2009arXiv0911.3200L}, and New Worlds \citep{2009SPIE.7436E...5C}, could measure the planet spectrum (in the visible, near infrared, or mid-infrared, depending on the technique), and its variability. Variations in the broadband mid-IR flux with the orbital phase, have been suggested in the framework of Darwin/TPF-I  as a way to identify planets with dense atmospheres \citep{2004ASPC..321..170S}, a necessary criterion for surface habitability, or to constrain obliquity  \citep{2004NewA...10...67G}. Variations of the reflected light \citep{2008ApJ...676.1319P} or the infrared emission \citep{2011Illeana} could reveal cloud patterns and the rotation rate of the planet. Until these large observatories are developed, we will have to study terrestrial exoplanets by observing the combined star+planet light. Although the absolute planetary flux cannot be extracted from such observations, the modulation of the flux at the orbital period of the planet can be attributed to the planet (assuming no pollution of the signal by intrinsic stellar variability or systematic instrumental effects). The phase curve of a hot Jupiter has already been detected in infrared bands in a nontransiting configuration \citep{2007MNRAS.379..641C,2010ApJ...723.1436C} and not detecting of the phase curve of the $\sim 7$~M$_{\oplus}$ planet GJ876d has been used to discuss the presence of a dense atmosphere attenuating the day/night difference in brightness temperature \citep{2009ApJ...703.1884S}. The phase- and wavelength-dependent emission of Hot Jupiters have been simulated using either 1D models assuming a local radiative-convective and chemical equilibrium \citep{2005ApJ...632.1132B} or circulation models using a Newtonian scheme for the temperature calculation and either equilibrium chemistry \citep{2010ApJ...719..341B} or kinetic calculations \citep{2006ApJ...652..746F}. The phase-dependent visible flux and its dependence on inclination has been studied for nontransiting by \citep{2011arXiv1101.1087K}. \citet{2011ApJ...726...82C} studied the thermal phase curve of hot Jupiters, including eccentric ones,  with a semi-analytic model.\\
In the present article we address the characterization of the atmosphere of a hot terrestrial planet that does not transit its host star, using spatially-unresolved spectro-photometry of the system. We study how the apparent emission of the planet varies with wavelength and orbital phase, for different atmospheric pressures, in the case of a tidally-locked planet. In another study \citep{2011maurin}, we discuss the retrieval of the radius, albedo, and inclination of airless planets. In this article we focus on the influence of the atmosphere, as simulated with a 3D climate model.

\section{Model}
\begin{table*}
\centering 
\begin{tabular}{|l|lllllllll|l|} 
\hline 
\hline 
name &   \tiny{HD40307b}$^{(a)}$ &      \tiny{GJ581e}$^{(b)}$ &      \tiny{GJ581b}$^{(b)}$ &      \tiny{GJ176b}$^{(c)}$ &     \tiny{61Virb}$^{(d)}$ &      \tiny{GJ876d}$^{(e)}$ &     \tiny{GJ1214b}$^{(f)}$ &      \tiny{GJ674b}$^{(g)}$ &     \tiny{55Cnce}$^{(h)}$ & \textbf{model} \\
\hline 
 $M$ [$M_{\oplus}$] &     4.77 &     2.19 &    18.48 &     9.70 &     5.88 &     7.85 &     5.7 &    15.01 &     8.57 & \textbf{9.5} \\
$a$ [au]  &    0.047 &    0.028 &    0.041 &    0.066 &    0.050 &    0.020 &    0.014 &    0.040 &    0.016 & \textbf{0.05} \\
$P$ [days]  &     4.31 &     3.15 &     5.37 &     8.80 &     4.22 &     1.93 &     1.58 &     4.69 &     0.73 & \textbf{8} \\
 $e$ &     0 &     0 &     0 &     0 &     0.12 &     0.21 &     0.27 &     0.20 &     0 & \textbf{0} \\
 $M_{*} [M_{\odot}]$ &     0.80 &     0.31 &     0.31 &     0.5 &     0.95 &     0.33 &     0.16 &     0.35 &     0.96 & \textbf{0.31} \\
 $d$ [pc] &     13.0 &      6.2 &      6.2 &      9.4 &      8.5 &      4.7 &     13.0 &      4.5 &     13.0 & \textbf{10} \\
\hline 
\hline 
\end{tabular}
\caption{Properties of known nearby low-mass planets ($d< 15$~pc, $M\sin{i} < 16 M_{\oplus}$) compared to the modeled planet. Except for GJ1214~b and 55Cnc~e, which are transiting, the mass is given for an assumed inclination of 60$^{\circ}$. Refs: $^{(a)}$\citet{Mayor2009b},$^{(b)}$\citet{Mayor2009a}, $^{(c)}$\citet{2009A&A...493..645F}, $^{(d)}$ \citet{Vogt2010}, $^{(e)}$ \citet{Rivera2010}, $^{(f)}$ \citet{2009Natur.462..891C}, $^{(g)}$ \citet{Bonfils2007}, $^{(h)}$ \citet{2011arXiv1104.5230W}.}
\label{knownPlanets} 
\end{table*}                   
 \begin{figure*}
\centering      
\includegraphics[width=0.8\linewidth]{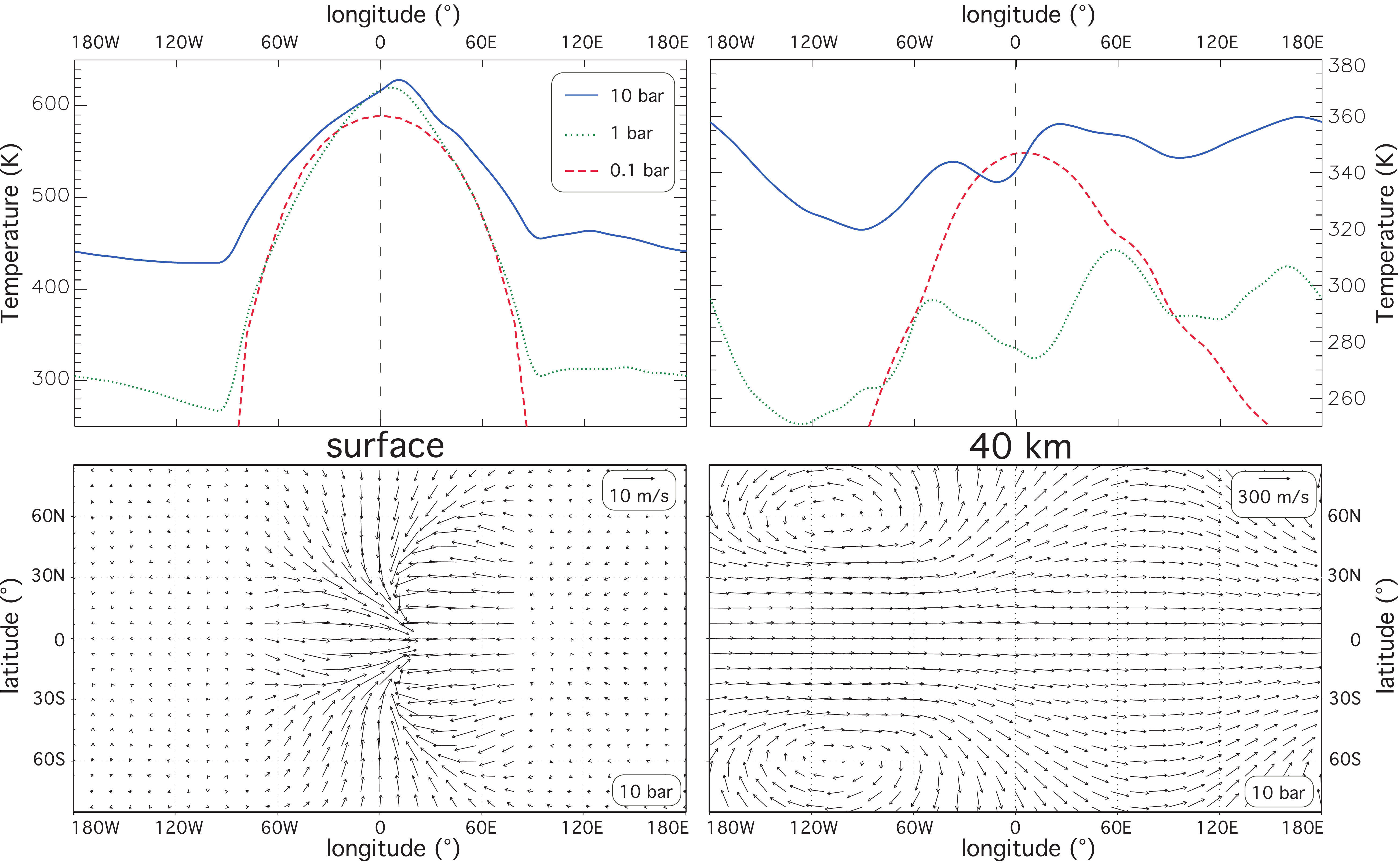}
 \caption{Equatorial temperature (top) and horizontal wind maps (bottom) at the surface (left) and 40 km (right). Temperatures are given for 0.1, 1 and 10 bar ; wind maps are shown for the 10 bar case only. The displacement of the hot spot, compared to the substellar point (longitude=$0^{\circ}$, latitude=$0^{\circ}$) can be seen on the surface plots.}
      \label{fig:circulation}%
    \end{figure*}
Modeling the orbital phase variations of the thermal emission of a planet observed by a distant observer requires us to model the 3D structure of the atmosphere, including the temperature, pressure, chemical composition, and cloud / aerosol content. Due to the computation time of 3D atmospheric simulations, the possible compositional diversity of planetary atmospheres, and the variety of planetary systems, this preliminary study does not aim to explore the effect of all the parameters controlling the observables (star type, orbital elements, planet rotation, planet size and gravity, atmospheric elemental composition, nature of the surface). At this stage, we chose to consider a simple planet corresponding to a system that we can expect to detect (for instance by radial velocity) in the vicinity of the Sun: a short-period massive terrestrial planet (sometimes called a \textit{hot super-Earth}) around an M dwarf. The planet we model may be representative of known nearby low-mass exoplanets, which we list in Table~\ref{knownPlanets}. \\
Our example planet is on a circular, synchronized orbit, with a null obliquity, a state consistent with strong tidal interactions (see section~\ref{subsec:nonsync}). We consider a fixed atmospheric composition and only vary the atmospheric pressure, in order to study how the atmosphere modifies the phase- and wavelength-dependent properties of the apparent infrared emission. The characteristics of the planetary system considered are summarized here: 

 \noindent \textbf{The star} is an M3 dwarf (0.31~M$_{\sun}$) with a luminosity of 0.0135~L$_{\sun}$ and the same spectral distribution as AD Leo (we used the Virtual Planet Laboratory AD Leo data from \citealp{Segura2005}). For flux and photon-noise calculations, the distance of the star is set to 10~pc.
 
 \noindent \textbf{The orbit} is circular with a semi-major axis of 0.0535 AU and period of 8 days. As seen on Fig.~\ref{fig:orbit}, the inclination of the orbit on the line of sight is set to 60$^{\circ}$, which the median value for randomly oriented systems. 
 
 \noindent \textbf{The planet} is a 1.8 R$_{\oplus}$ and 9.5 M$_{\oplus}$ terrestrial planet, consistent with an Earth-like bulk composition \citep{2006Icar..181..545V,2007Icar..191..337S}. Its surface gravity is 3 $g$. We set the surface albedo to 0.2. The planet rotation period is assumed to be equal to its orbital period (8~days).
 
  \noindent \textbf{The atmosphere} is made of carbon dioxide (CO$_2$) as the unique constituent. We consider three atmospheric pressures: 0.1, 1, and 10 bars, as well as a case with no atmosphere. 

In the case with no atmosphere, the surface temperature is in local equilibrium with the stellar irradiation and given by 
\begin{equation}
T_{\rm eq}(\theta) = [S(1-A) \cos(\theta)/\sigma]^\frac{1}{4} 
\end{equation}
where $S$ is the stellar flux at 0.05~AU, $A$ the surface albedo, $\sigma$ the Stefan-Boltzmann constant and $\theta$ the zenith angle. The emissivity of the surface is set to 1. On the dark hemisphere, the temperature is set to 37~K. This value is obtained by scaling the intrinsic luminosity of the Earth ($\sim 30$~TW, due mainly to radioactive decay) to the mass and radius of the planet. Using a night-side temperature of 0~K would not, however, change the disk-integrated properties in an observable way. The substellar temperature is 553~K,  the equilibrium temperature is 390~K (stellar irradiation uniformly distributed over the whole planet), and the equilibrium temperature calculated only for the starlit hemisphere is 465~K ( stellar irradiation uniformly distributed over the starlit hemisphere only). We do not consider unsynchronized or eccentric cases, and so the irradiation at a given location on the planet always remains the same. Therefore we do not need to include the effects of surface thermal inertia. \\
\begin{figure*}[p]
\centering
\includegraphics[width=\linewidth]{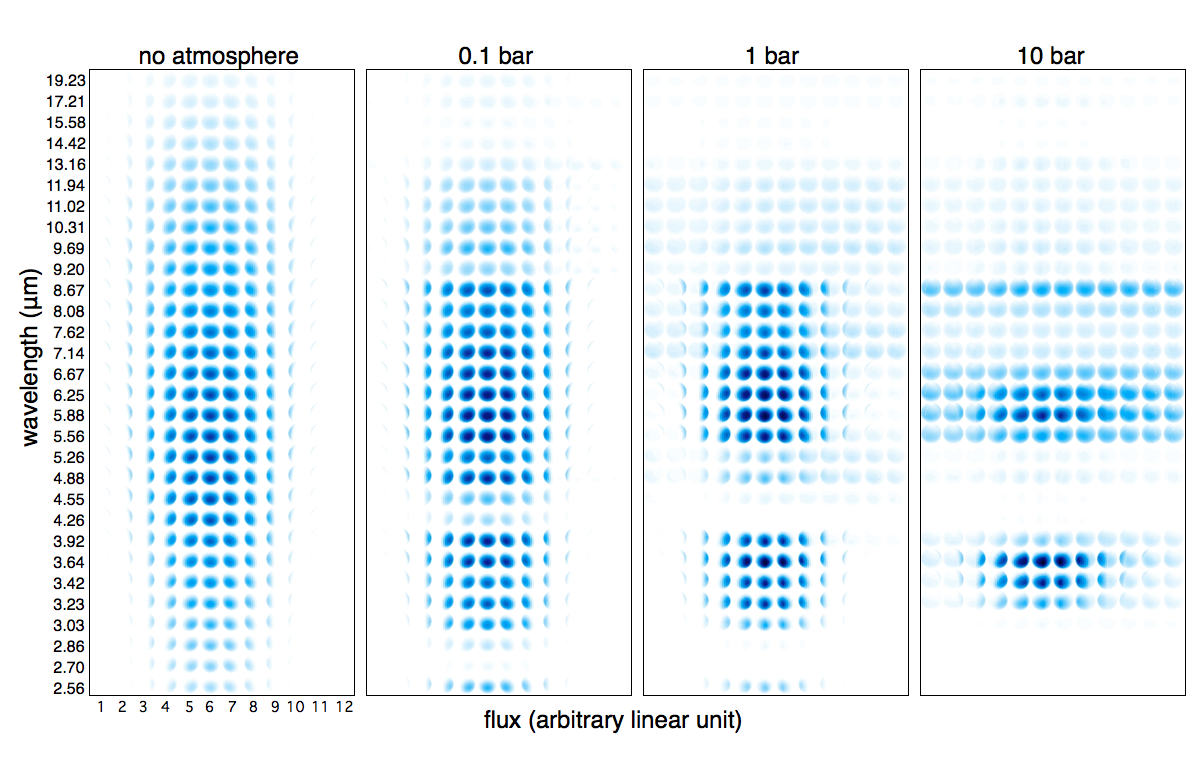}
 \caption{Planetary emission maps as a function of wavelength and phase. Phase numbers refer to Fig.~\ref{fig:orbit}.}
      \label{fig:mapflux}%
\includegraphics[width=\linewidth]{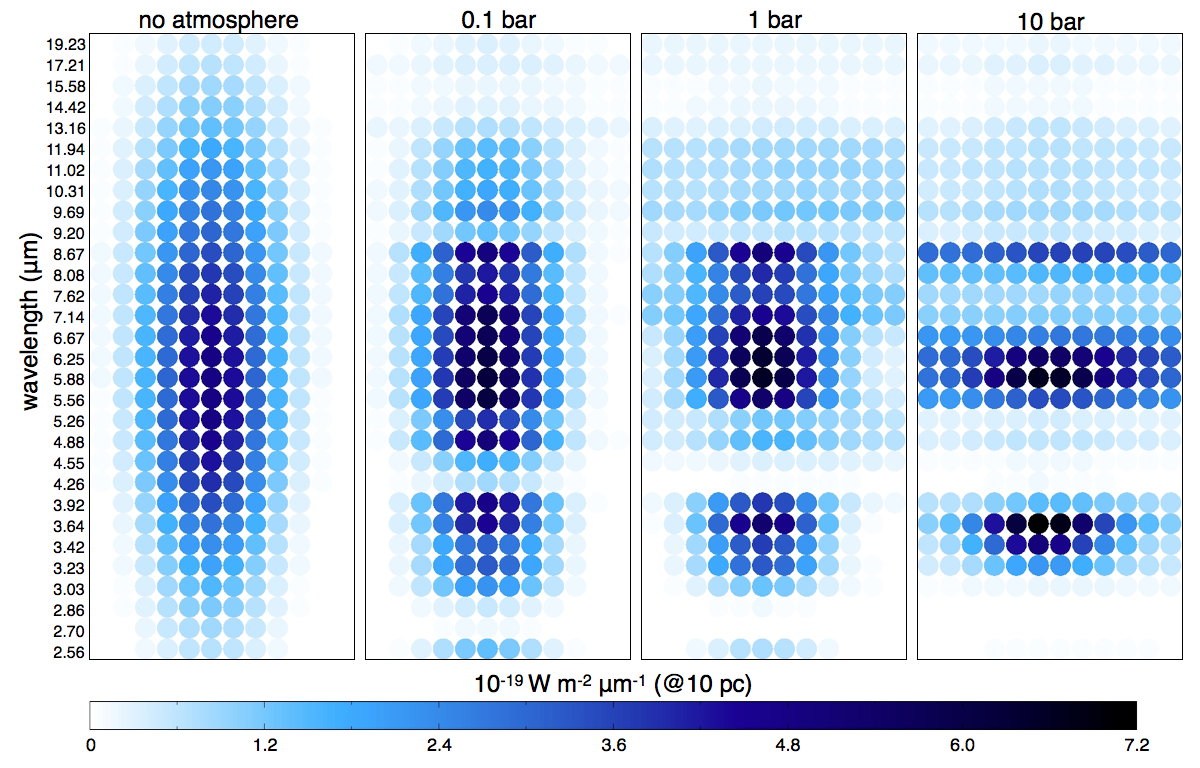}
 \caption{Disk-integrated flux as a function of wavelength and phase}
      \label{fig:diskflux}%
     \end{figure*}

For the three cases with an atmosphere, we used the LMD 3D atmospheric general circulation model (GCM).
This model, originally developed for the Earth \citep{Sadourny1975, Hour:06}, has been adapted to most terrestrial
planet atmospheres in the solar system like Mars \citep{Forget1999}, Titan \cite[]{Hour:95}, Venus \citep{Lebo:09} and
even Triton \citep{Vang:10}. Because GCMs are based on physical
equations that apply across wide ranges of pressure, temperature, and atmospheric composition, they can simulate accurately the thermal structure and winds of a broad diversity of atmospheres, provided that basic data such as spectroscopic properties are available in the regime of interest. In this study, we used a new, generic version of the LMD GCM that has been developed
specifically for exoplanet and paleoclimate studies. It has already been used to study the early Mars climate
\citep{2010DPS....42.4502F} and possible climates on GJ581d \citep{Wordsworth2011}. In practice, this version uses 
the LMDZ 3D dynamical core \citep{Hour:06}. It is based on a finite-difference
formulation of the classical primitive equations of meteorology. The primitive equations are a simplified
version of the Navier-Stokes general equations of hydrodynamics based on three main approximations: 
the atmosphere is assumed 1) to be a perfect gas, 2) to remain in hydrostatic
equilibrium vertically, and  3) to have much
smaller vertical dimensions than the radius of the planet (thin-layer approximation). In the LMDZ GCM, 
the potential enstrophy and total angular momentum
are numerically conserved for barotropic flows \citep{Sadourny1975}.
Scale-selective hyperdiffusion was used in the horizontal plane for stability, and linear 
damping was applied in the topmost levels to eliminate the artificial reflection of vertically propagating waves. 
The planetary boundary layer is parameterized using implicit timestepping and the method of Mellor \& Yamada 
\citep{Mellor1982,Galperin1988} is used to calculate turbulent mixing.  In addition to this  parametrization, a convective
adjustment scheme is used to prevent subadiabatic vertical temperatures
gradients. The temperature of the surface is computed
from  the radiative, sensible and latent heat fluxes at the surface
using an 18-level model of thermal diffusion in the soil and assuming a homogeneous thermal inertia of
250~J~m$^{-2}$~s$^{-1/2}$~K$^{-1}$ and a surface albedo of 0.2.  \\ 

GCMs have already been applied to terrestrial exoplanets, especially to study their habitability when locked in synchronous rotation \citep{1997Icar..129..450J, 2010JAMES...2...13M,2011ApJ...726L...8P,2011MNRAS.tmp..370H}. The main improvement of the model we use is the radiative transfer scheme, which is similar to the one described in \cite{Wordsworth2010b}. We computed high-resolution spectra over 
a range of temperatures and CO$_2$ pressures using a 14 $\times$ 9 temperature-pressure grid with values $T 
= 100-750$~K, $p = 10^{-3}-10^5$~mbar. The correlated-$k$ method is then employed to produce a smaller database of coefficients 
suitable for fast calculation in a GCM. The model uses 36 bands between 0.3 and 5~$\mu$m for the incoming stellar radiation 
and 38 bands above 2.5~$\mu$m for the planetary infrared emission. The overlap is due to the fact that no source function 
is considered for the stellar light transfer, only absorption and scattering. Sixteen points were used for the $g$-space 
integration, where $g$ is the cumulated distribution function of the absorption data for each band. CO$_2$ 
collision-induced absorption (CIA) was included using a parameterization based on the most recent theoretical and experimental 
studies \citep{Wordsworth2010,Gruszka1997,Baranov2004}.  A two-stream scheme \citep{Toon1989} was used to account for the 
radiative effects of Rayleigh scattering, which was included by the method described in \citet{Hansen1974}. Condensation 
and CO$_2$ clouds were not included in these simulations, which is consistent with the range of temperature and pressure 
found in the 1 and 10~bars cases but not in the 0.1~bar case (see section~\ref{subsec:condense}).
%
 \begin{figure*}
\centering
\includegraphics[width=\linewidth]{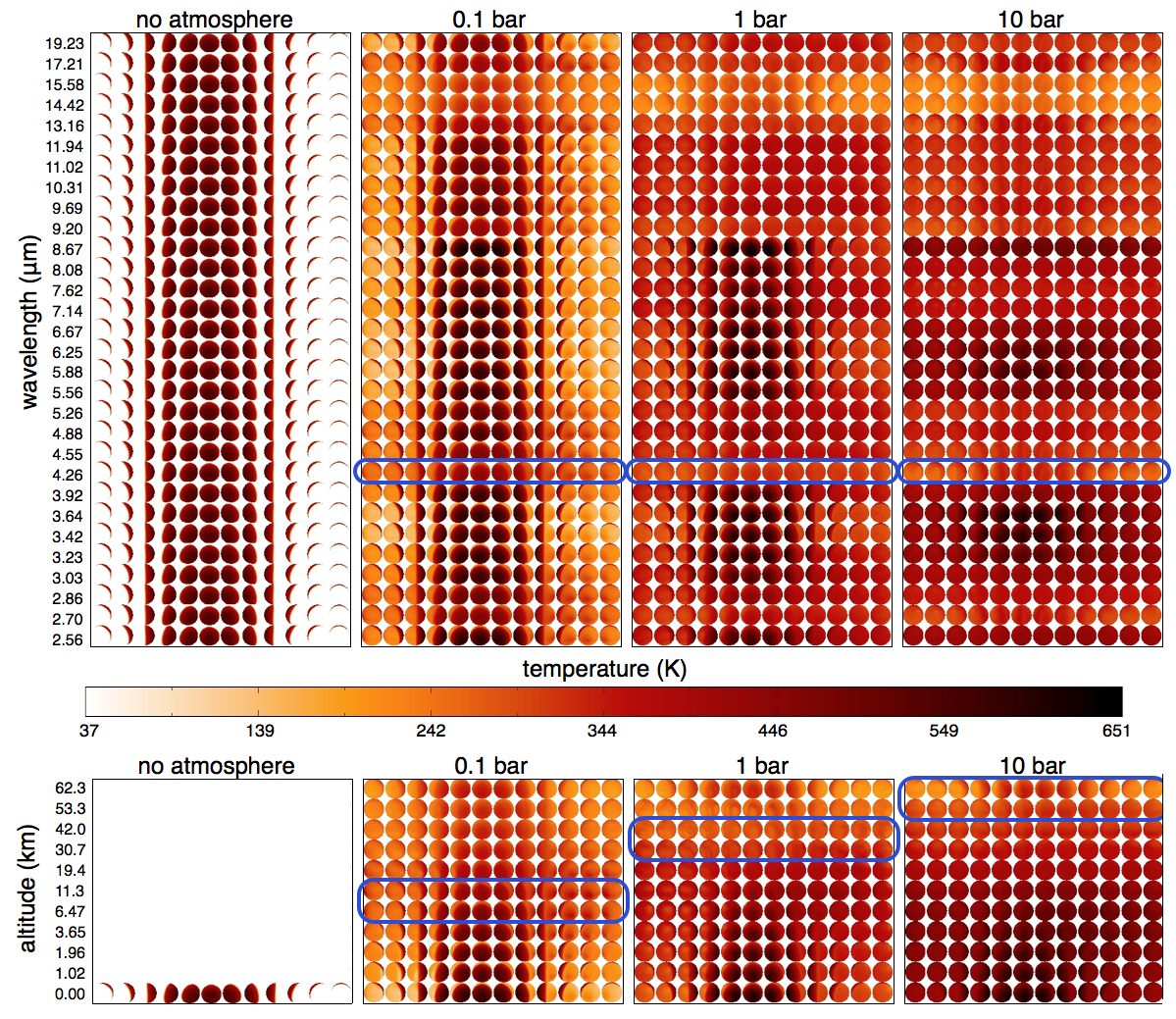}
 \caption{Brightness temperature maps as a function of wavelength and phase (top) and gas temperature maps as a function of altitude and phase (bottom).}
      \label{fig:tbright}%
    \end{figure*}

The modeled atmospheres were made of a single atmospheric constituent (CO$_2$). The GCM could include more species and real atmospheres are not thought to consist of only one constituent, but this has several advantages for a preliminary study. A key point of this study is to show that the variation spectrum (peak amplitude of the phase curves as a function of wavelength), which can be measured in relative photometry for a nontransiting planet, is strongly correlated with the emission spectrum and exhibits molecular signatures. We therefore preferred to study a single component atmosphere as a first step, so that the complex relation between the composition, the spectral- and spatial distribution of the emission, and the amplitude spectrum of the phase curves could be understood more easily. This pure CO$_2$ case can be seen as an illustration of the principles of this characterization method. Future, more realistic studies will include atmospheric mixtures and, in particular, water vapor and clouds. \\
\begin{table}[tb]
\caption{The 30 infrared bands used to compute phase curves ($\mu$m)}              
\label{tab:bands}      
\begin{tabular}{c c c | c c c}          
\hline\hline                        
$\lambda_{center}$ & $\lambda_{min}$& $\lambda_{max}$\ & $\lambda_{center}$ & $\lambda_{min}$& $\lambda_{max}$\\    
\hline                                   
   2.56 &    2.53 &    2.63 &    6.67 &    6.46 &    6.90 \\
   2.70 &    2.63 &    2.78 &    7.14 &    6.90 &    7.38 \\
   2.86 &    2.78 &    2.94 &    7.62 &    7.38 &    7.85 \\
   3.03 &    2.94 &    3.13 &    8.08 &    7.85 &    8.37 \\
   3.23 &    3.13 &    3.32 &    8.67 &    8.37 &    8.93 \\
   3.42 &    3.32 &    3.53 &    9.20 &    8.93 &    9.44 \\
   3.64 &    3.53 &    3.78 &    9.69 &    9.44 &   10.00 \\
   3.92 &    3.78 &    4.09 &   10.31 &   10.00 &   10.66 \\
   4.26 &    4.09 &    4.40 &   11.02 &   10.66 &   11.48 \\
   4.55 &    4.40 &    4.71 &   11.94 &   11.48 &   12.55 \\
   4.88 &    4.71 &    5.07 &   13.16 &   12.55 &   13.79 \\
   5.26 &    5.07 &    5.41 &   14.42 &   13.79 &   15.00 \\
   5.56 &    5.41 &    5.72 &   15.58 &   15.00 &   16.39 \\
   5.88 &    5.72 &    6.07 &   17.21 &   16.39 &   18.22 \\
   6.25 &    6.07 &    6.46 &   19.23 &   18.22 &   20.31 \\
 \hline                                             
\end{tabular}
\end{table}  
For this study the GCM resolution is $32 \times 24 \times 16$ grid (longitude $\times$ latitude $\times$ altitude). 
This is found to be high enough to capture the most important features of the atmospheric dynamics, 
given that it resolves the Rossby deformation radius $L_R \sim \sqrt{R T} \slash 2\Omega \sim 2.5 R_{\oplus}$, where $R$ is the specific gas constant, $\Omega$ is the planetary rotation rate, and $T$ is the atmospheric temperature. We did not test the effect of our model's finite difference numerical scheme on the results. However, this issue has been studied by Heng et al. \citeyearpar{2011MNRAS.tmp..370H}, who found quantitative agreement between spectral and finite difference dynamical cores for simulations of a hypothetical tidally locked Earth.\\
The model starts with an isothermal surface and atmosphere at 300~K. It computes the evolution of the atmosphere during 100 orbits, after which the state of the atmosphere is close enough to a steady state. Using the last orbit, we produce phase curves for 30 of the 38 infrared bands (given in Table~\ref{tab:bands}), between 2.5 and 20~$\mu$m. To do that, we use the top-of-the-atmosphere longitude-latitude maps of outgoing fluxes computed by the GCM with a timestep of 1 hour (192 points on the orbit). At a given time and wavelength, we sum all these outgoing fluxes, weighted by the area of their corresponding cell and by the cosine of the angle between the normal to the surface and the direction towards the observer. We assume an isotropic distribution of specific intensities at the top of the atmosphere (see section~\ref{subsec:lambert} for a discussion on the validity of this approximation). For the case with no-atmosphere, the surface emission is calculated using the same longitude-latitude cells and the blackbody emission is integrated over the same spectral bands. The procedure for deriving the phase curve is then exactly the same.
\section{Results}
The results of our model are presented for initial pressures of 0, 0.1, 1 and 10~bar. In Figure~\ref{fig:circulation}, we present some characteristics of the atmospheric circulation and associated heat transport we have obtained with the model. Wind maps (given in the 10~bar case) show that surface winds converge toward the substellar area, feeding the convective plume generated by the stellar heating. Coriolis forces 
due to the planet's rotation
produce an east-west asymmetry in the wind field. In the upper atmosphere, super-rotation occurs with an equatorial jet slightly faster than the planet rotation itself. In the uppermost layer, wind speeds become close to the speed of sound, and we may be approaching the limit of our model validity. However, this has negligible consequences on the dynamics at higher pressures that produce the thermal emission. The super-rotation regime starts for pressures above about 1~bar and is not found in the 0.1 bar case. 

In the surface wind map for the 10~bar case, the convective plume does not start at the exact substellar location (longitude=0$^{\circ}$, latitude=0$^{\circ}$) but is slightly shifted eastward. This shift of the hot spot is due to heat transport by the atmosphere. It has been predicted for Hot Jupiters \citep{2002A&A...385..166S} and was observed with Spitzer \citep{2007Natur.447..183K} as a lag between the maximum of the phase curve and the  secondary eclipse. This delay is also found in our model depending on the wavelength and the probed altitude. Figure~\ref{fig:circulation} also shows the equatorial temperature as a function of longitude. On the surface, a 5 and a 10$^{\circ}$ displacement of the hot spot can be seen for the 1 and 10~bar atmospheres, respectively, corresponding to a 2.5 and a 5~hour lag. However, at 40~km and for the 1 and 10~bar cases, the longitudinal thermal profile is much more complex and no main maximum can be identified. The existence of a clear maximum and its associated lag thus depends strongly upon the wavelength. If observed in a single band, a lag can be regarded as evidence of an atmosphere only if the planet is definitely synchronized. Indeed, surface thermal inertia can produce a similar hot spot displacement if the planet is not synchronized. In particular, the case of eccentric planets that can be either in a pseudo-rotation state or in various spin-orbit resonances \citep{Leconte2010} must be treated with caution. \\

How efficiently the heat deposited by insolation is redistributed to the night side and to the poles can be evaluated by comparing analytical radiative and dynamical timescales. The mean radiative timescale $\tau_{R}  = \frac{c_{P}P} {\sigma g T^{3}}$ where $c_{P}$ is the heat capacity of CO$_2$ (assumed constant here following \citealp{Wordsworth2010b}) and T is the mean bolometric brightness temperature of the planet ($\sim 400$~K in all cases) . The mean dynamical timescale $\tau_{D}=R/\bar{v}$ where R is the planetary radius and $\bar{v}$ is the density-weighted average of the wind speed. The ratio $\tau_{R}/\tau_{D}$ is $\sim 0.1$ for $P=0.1$~bar, $\sim 1$ for $P=1$~bar, and $> 5$  for $P=10$~bar. Heat redistribution is thus efficient for pressures above 1~bar. \\
 \begin{figure*}
\centering
\includegraphics[width=0.75\linewidth]{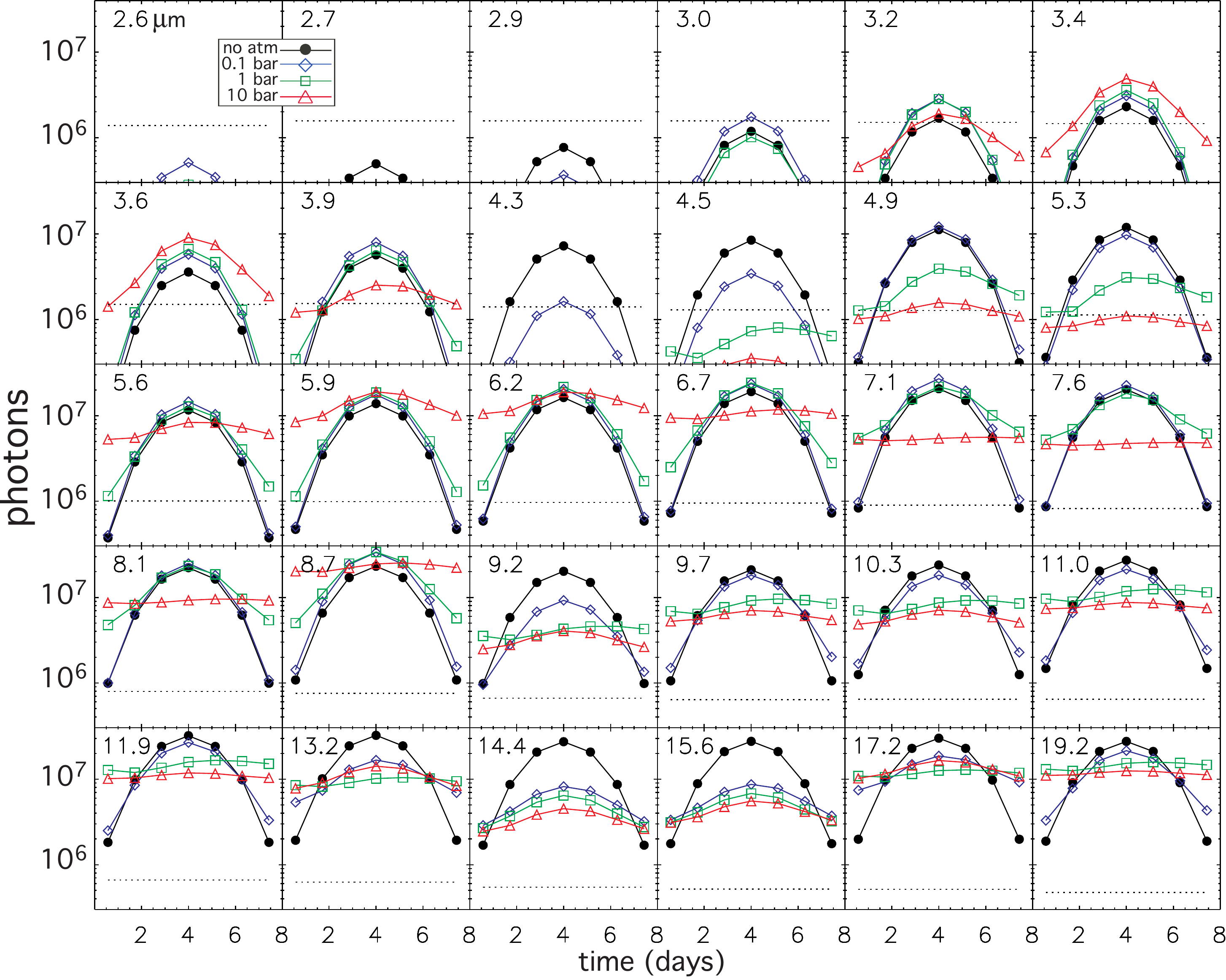}
 \caption{Planetary signal as a function of time. The 8-day orbits is divided in 7 exposures. The number of photons is given per exposure, per band, for a 6m telescope and a 10~pc target. The dotted line represents the amplitude of the stellar photon noise. Wavelengths are indicated in the upper left corner of each plot.}
      \label{fig:phaseJWST}%
       \end{figure*}
\begin{figure*}
\centering
 \includegraphics[width=0.75\linewidth]{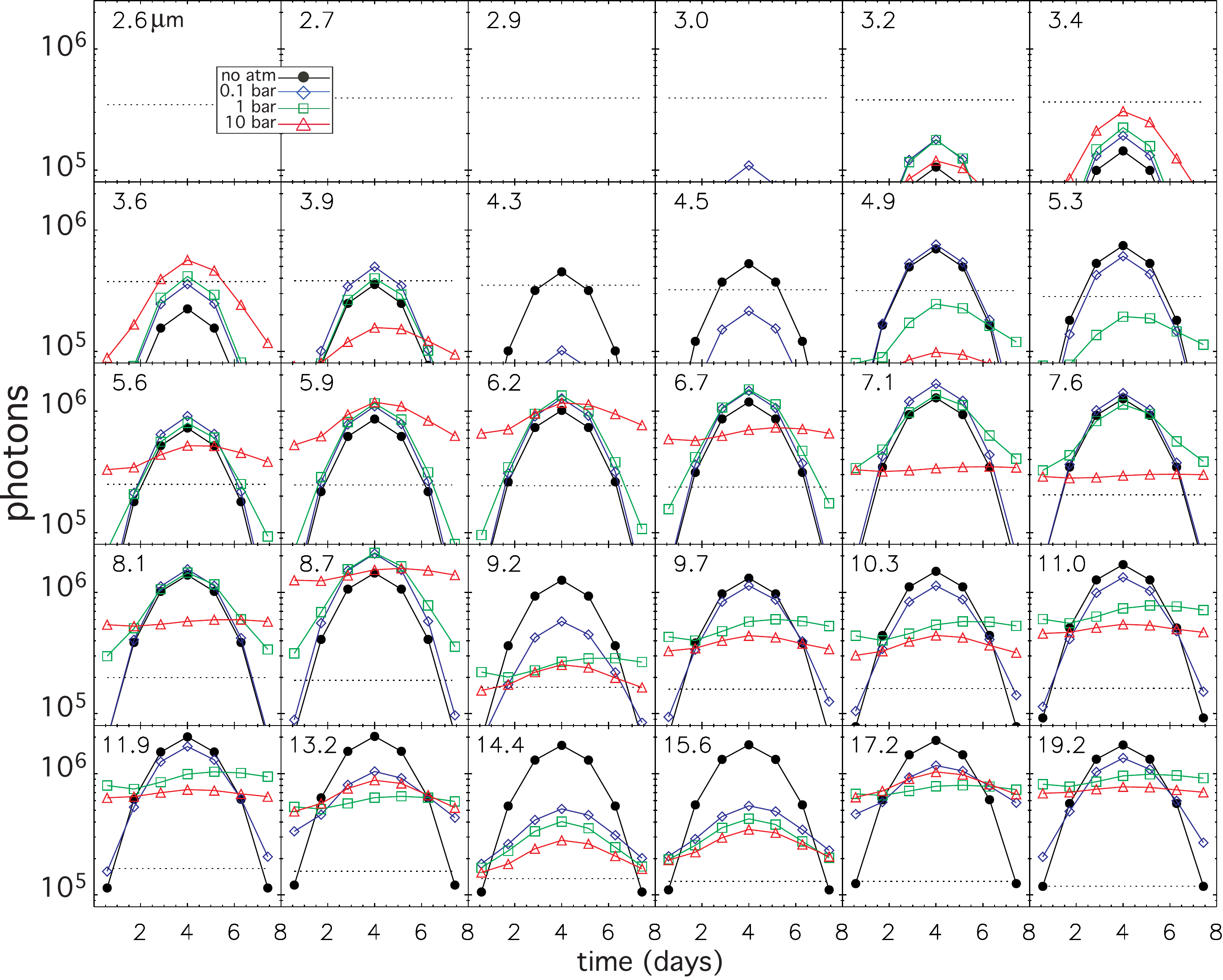}
 \caption{Same as Fig.\ref{fig:phaseJWST} but for a 1.5~m telescope.}
      \label{fig:phaseEChO}%
 \end{figure*}
 \begin{figure*}
\centering      
 \includegraphics[width=0.75\linewidth]{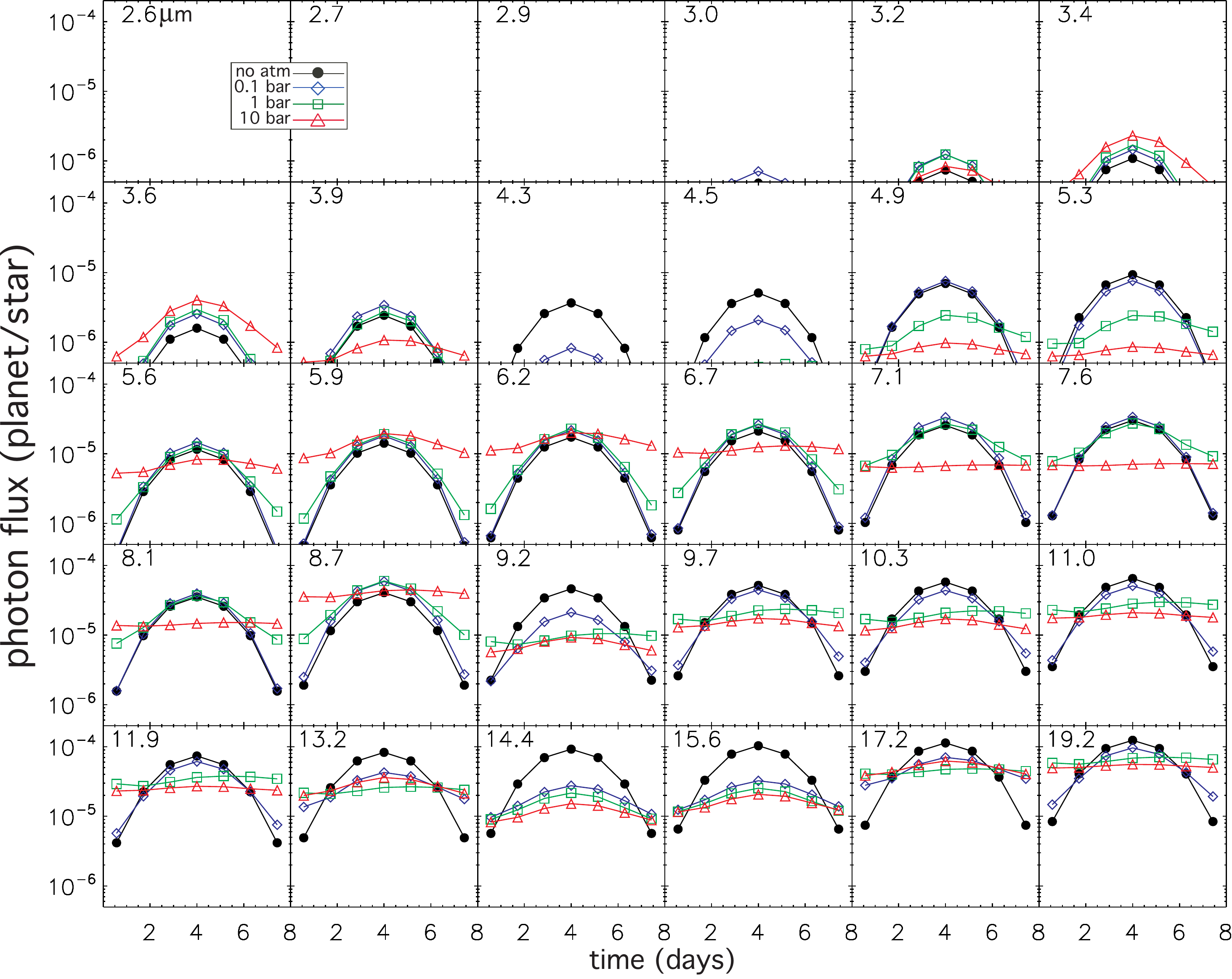}
 \caption{Same as Figs.~\ref{fig:phaseJWST} and \ref{fig:phaseEChO} but the photon flux is given here as the planet/star contrast ratio.}
      \label{fig:phase_contrast}%
          \end{figure*}
          
Figures~\ref{fig:mapflux} and \ref{fig:diskflux} show the top-of-the-atmosphere outgoing fluxes, as seen by the observer, as a function of wavelength (Y-axis) and phase (X-Axis). The phases represented and their numbering are the same as in Fig.~\ref{fig:orbit}. Fig.~\ref{fig:mapflux} gives the spatial distribution of thermal emission (in arbitrary units proportional to W m$^{-2} \mu$m$^{-1}$). The aim of this figure is to show the spatial distribution of the emission on the apparent planetary disk. Figure~\ref{fig:diskflux} gives the corresponding disk-integrated fluxes calculated at 10~pc. The bond albedo of the planet is only slightly affected by the atmospheric pressure:  its value is 0.20, 0.22, 0.21, and 0.205 for 0, 0.1, 1, and 10~bars, respectively. Therefore, the global thermal emission of the planet in Watts does not change between the different cases, but its spectral and spatial distribution strongly depends on the pressure due to the effects of the atmosphere on the radiative and heat transport. With increasing pressure, the planet tends to radiate predominantly in the spectral atmospheric windows (where CO$_2$ absorbs less), while the emission in the dark hemisphere becomes more important due to the more efficient heat transport through atmospheric circulation. 
    
 \begin{figure*}
\centering      
\includegraphics[width=0.75\linewidth]{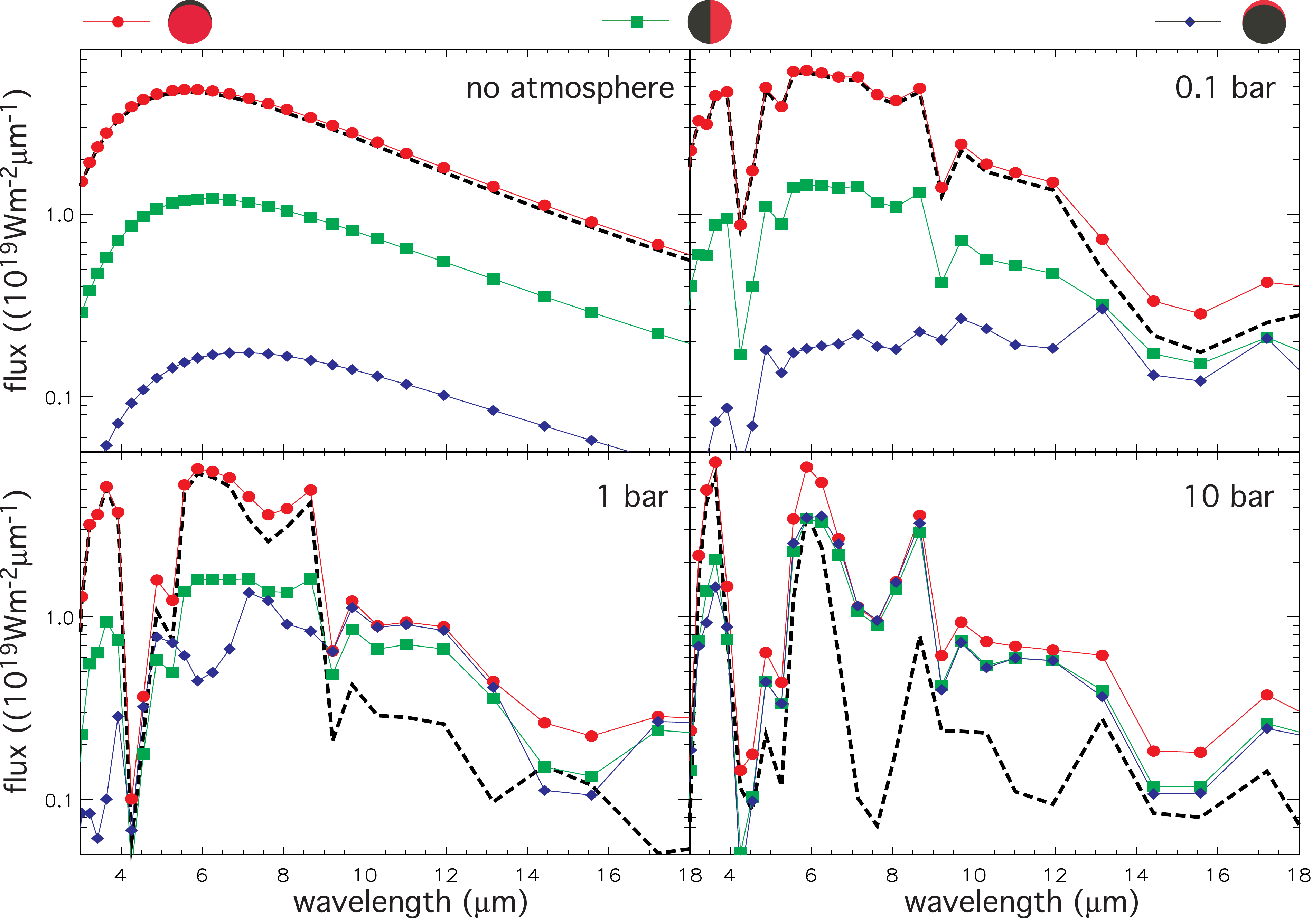}
 \caption{Planetary spectrum vs. variation spectrum. Circles, squares, and diamonds correspond the planetary spectrum at 3 different phases indicated at the top of the graph (phases 6, 3, and 12 in Fig.~\ref{fig:orbit}). The dashed line shows the amplitude of the 7-point phase curves as a function of wavelength (i.e. variation spectrum).}
      \label{fig:var_spectrum}%
    \end{figure*}
  At 0.1~bar, the absorption features of CO$_2$, in particular the 2.7, 4.3, and 15~$\mu$m bands, already affect the spectral distribution, but the thermal emission on the night side remains negligible. In atmospheric windows, for instance between 5 and 8~$\mu$m, the flux variations do not differ much from the airless case, but there is slightly higher maximum emission, as surface temperatures are increased by greenhouse warming. At higher pressures, weaker CO$_2$ bands (for instance between 9 and 11~$\mu$m) and collision-induced absorption (dimer absorption at 7-8~$\mu$m) also shape the emission spectrum. Night-side cooling is noticeable at 1 bar and becomes a significant contribution to the total emission at 10~bars.
Figure~\ref{fig:tbright} presents flux maps similar to those in Fig.~\ref{fig:mapflux} but converted into brightness temperatures $T_B$ (= temperature of a blackbody with the same radiance at the considered wavelength). At the bottom of Fig.~\ref{fig:tbright}, we have mapped the gas temperature at different altitudes on the observed planetary disk. With increasing pressure, the surface and the lower atmosphere become hotter owing to greenhouse warming and more uniform thanks to efficient heat transport (see Table~\ref{tab:temps}) but the atmosphere also becomes more opaque at a given wavelength. Therefore, the observed emission comes from higher altitudes. With the exception of some nightside regions of the 0.1~case, and some high-altitude substellar inversion in the 1 and 10~bar cases, temperature monotically decreases with increasing altitude. This allows us to roughly determine the altitude at which the photons of a given wavelength are emitted (the $\tau \sim 1$ transition region between the optically thick and thin atmospheric layers) without computing the contribution function, by comparing the maps of $T_B(\lambda)$ and $T(z)$. For instance, we can see that the 4.26~$\mu$m band (framed in Fig.~\ref{fig:tbright}) is emitted around 10~km, 35~km, and 60~km, in the 0.1, 1, and 10~bars simulations, respectively . Therefore, the increase in temperature in the lower atmosphere owing to enhanced CO$_2$ level results in higher fluxes only in atmospheric windows. Windows open at low pressure are closed at high pressures by the CIA continuum and weak features. Also, the day-night temperature contrast is more pronounced at the surface and in the lower atmosphere than at higher altitudes. As a consequence, wavelengths probing the lower atmosphere exhibit large phase-dependent variations, while wavelengths falling in opaque regions of the planet spectrum and probing higher levels exhibit small variations. This can be seen for instance at 8.7~$\mu$m, which probes the first few kms of the atmosphere in the 0.1 and 1~bar cases, exhibiting high fluxes and large variations. At 10~bar, the lower atmosphere becomes optically thick due to  CO$_2$-CO$_2$ dimer absorption and the flux emerges above 10~km with smaller values and variations.
\subsection{Observability of the phase curves}

The variation in the apparent emission of the planet produces a modulation of the observed star+planet combined flux. Assuming that the planet is detected by radial velocity, its orbital period is known. This allows the observer to extract the flux variations at this specific period. We do not address here the issues of stellar variability and instrumental stability that certainly represent the main obstacles in extracting the planet signature. As a first step, we simply compare the planet emission with the stellar-photon noise to check that the observation can at least be considered. To calculate the stellar photon noise, we need to assume a collecting area, an exposure time, a spectral resolution, and a target distance. In this study, the spectral resolution is imposed by the bandwidths of our radiative transfer scheme (band limits are given in Table~\ref{tab:bands}). The corresponding resolution varies from $\sim 25$ at 2.5~$\mu$m to $\sim 10$ at 20~$\mu$m. The target distance is set to 10~pc. We assume that one complete orbit of the planet is observed with an exposure time set to one seventh of the orbital period (27.4 hours). An odd number of exposures allows us to center one exposure on the maximum phase. \\
    
Figures~\ref{fig:phaseJWST} and \ref{fig:phaseEChO} show the multiband phase curves for telescope diameters of 6 and 1.5~m, respectively. These values correspond to the size of JWST\footnote{The collecting area of JWST is 25~m$^2$} and EChO, a space telescope proposed to ESA for Cosmic Vision 2. With a 6~m ideal collector, the signal to photon-noise ratio exceeds 5 for $\lambda > 3.5~\mu$m and 10 for $\lambda > 3.5~\mu$m, in the scenario giving the largest variations, which is the case with no atmosphere. With 1.5~m, SNRs of 5 and 10 are obtained at wavelengths above 4.5 and 8.5~$\mu$m, respectively. Figure~\ref{fig:phase_contrast} gives the same phase curves but the flux is given relative to the star. We can see that a photometric precision on the order of $10^{-5}$ is required to resolve the planetary signal, which is the aimed precision for EChO in the infrared and the precision reached by Kepler in the visible. \\
 \begin{figure}[htb]
 \centering
\includegraphics[width=\linewidth]{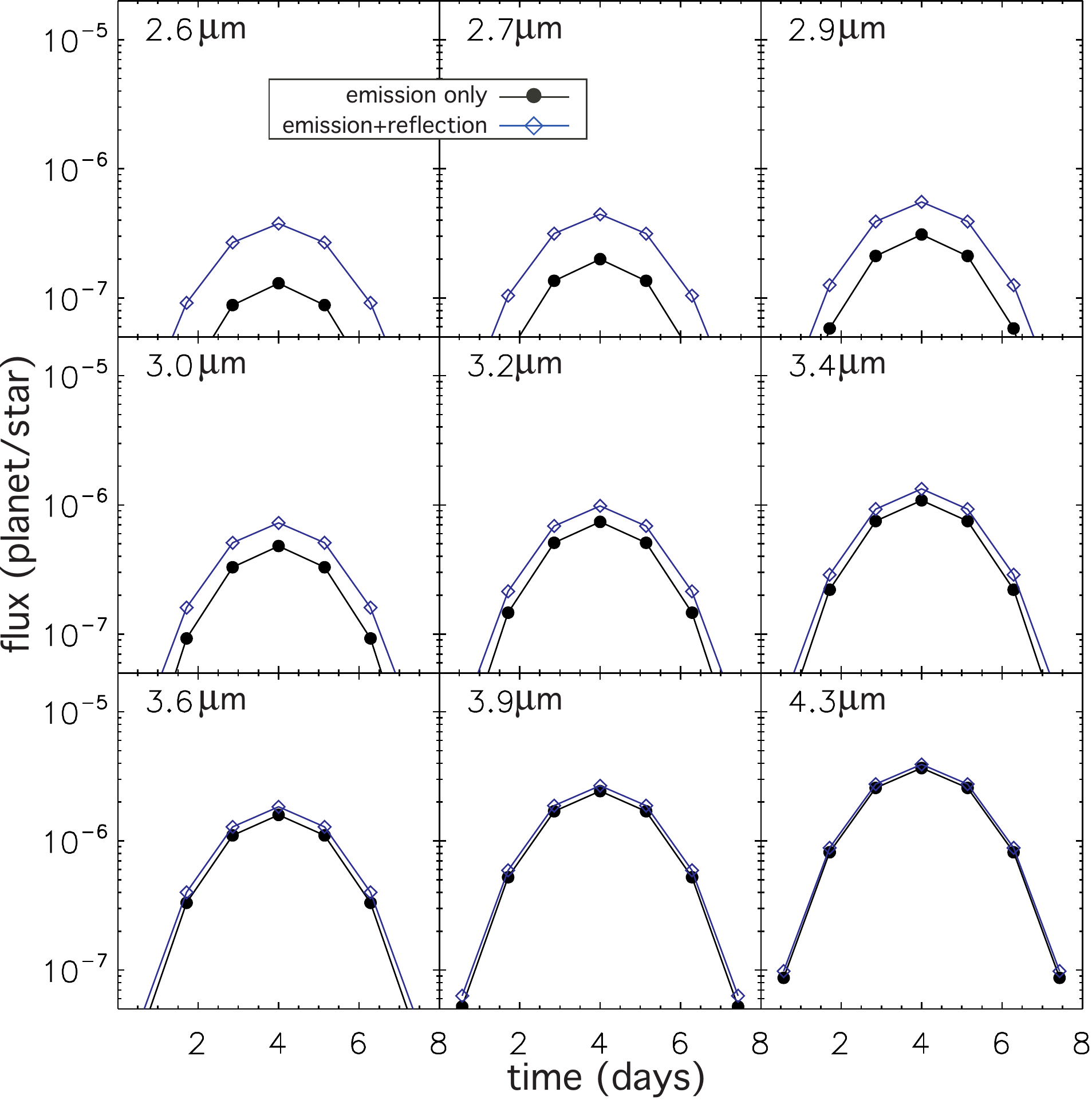}
 \caption{Contribution of the reflected light to the phase curve with no atmosphere. The 7-point phase curves are computed with emission only (black circles) and with emission+reflection (diamonds). Band central wavelengths are indicated in the upper left corner of each plot. }
      \label{fig:refl}%
    \end{figure}
The phase curves have been computed using the thermal emission given by the GCM and neglecting the stellar light scattered by the planet.  As shown in Fig.~\ref{fig:refl}, the contribution of the reflected light is only significant at wavelengths below 3.5~$\mu$m. At these short wavelengths, the planet/star ratio is lower than $10^{-6}$, making the observation of the planetary signal out of reach even for the next generation of infrared detectors. At higher wavelengths, where the extraction of the planetary emission is conceivable, we can thus safely neglect the reflected light.
\subsection{The variation spectrum}
Observation of the unresolved star+planet light does not give access to the absolute planetary emission (unless a secondary eclipse is observed). What can be measured is the variation in the planet emission. Valuable information about the nature of the atmosphere can be obtained from the \textbf{variation spectrum}: the peak amplitude of the phase-dependent variations as a function of the wavelength.  
    Figure~\ref{fig:var_spectrum} shows the variation spectrum compared to the planetary spectrum at three different phases. For the 0 and 0.1~bar cases, the mid-IR variation spectrum is extremely close to the emission spectrum at maximum phase. This is because the night side emission is extremely weak. In the 1~bar case, the variation spectrum and the maximum phase spectrum are similar between 3 and 9~$\mu$m. Above that, there is no atmospheric window probing the lowermost part of the atmosphere and the variations become very small. At 10 bars, 2 windows remain, at 3.5 and 6~$\mu$m, plus a marginal one at 9~$\mu$m. At other wavelengths, the planet spectrum is basically phase-independent. In all cases with an atmosphere, the variation spectrum is shaped by the radiative properties of CO$_2$, and the presence of CO$_2$ (or another IR absorber) can be inferred from the variation spectrum.

\begin{table}[htb]
\caption{Mean and extreme surface temperatures. }              
\label{tab:temps}      
\begin{tabular}{l l l l }          
\hline\hline                        
&T$_{\rm mean}$ (K) & T$_{\rm max}$ (K) & T$_{\rm min}$ (K) \\    
\hline                                   
no atmosphere &  234 & 553  &  37 \\
0.1 bar & 302 &  589 &  120$^{(a)}$,  170 $^{(b)}$ \\
1 bar &  370 & 621.5 & 227.5  \\
10 bar & 470 & 633 & 414  \\
\hline                                   
 \hline                                             
 \multicolumn{4}{l}{$^{(a)}$ {\tiny With CO$_2$ condensation turned off}} \\
  \multicolumn{4}{l}{$^{(b)}$ {\tiny CO$_2$ condensation temperature at 0.1 bar. }} \\
\end{tabular}
\end{table}

\section{Discussions}
\subsection{CO$_2$ condensation} \label{subsec:condense}
 \begin{figure}[hbt]
\centering
\includegraphics[width=\linewidth]{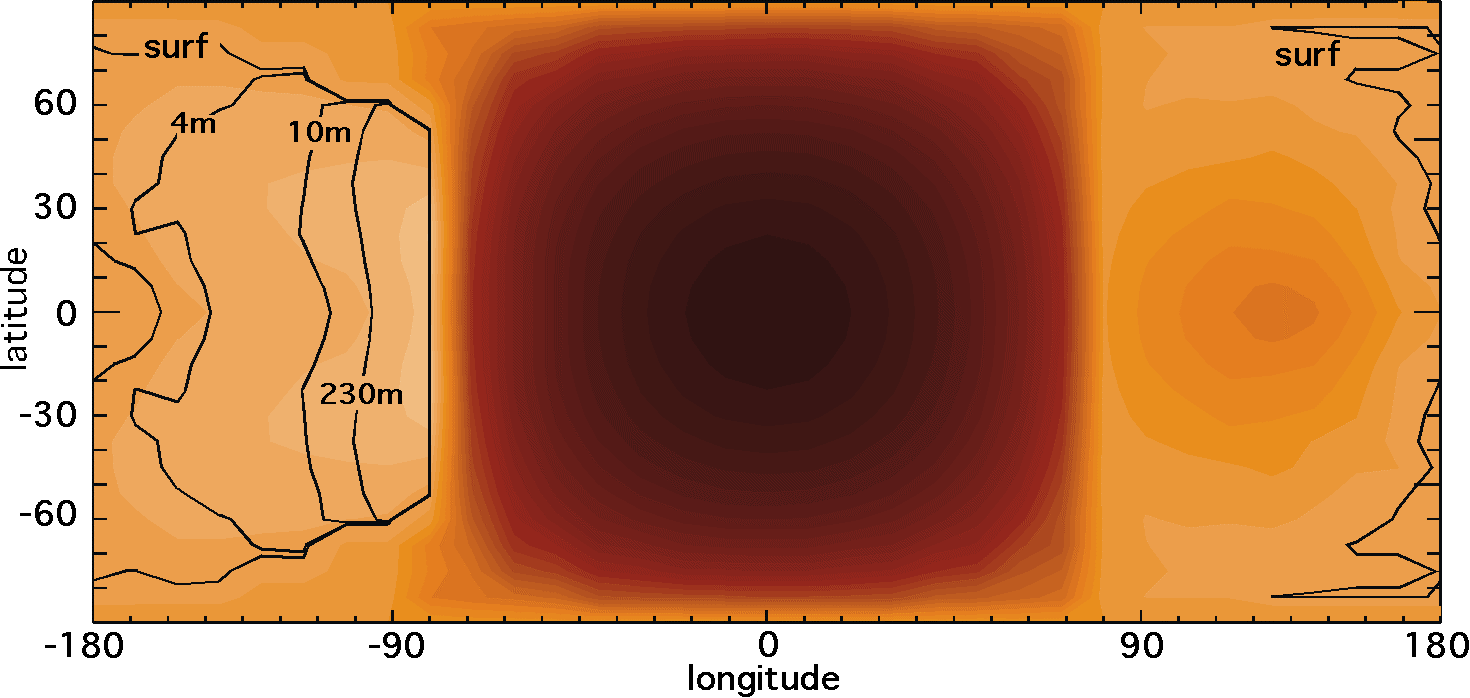}
 \caption{Condensation of CO$_2$ in the 0.1~bar case. Color contours indicate the surface temperature, with the same color scale as in previous temperature maps. Contour lines indicate the region where CO$_2$ is supersaturated at the surface, at 4~m, 10~m and 230~m. Levels above 230~m are not saturated. The $0^{\circ}$-$0^{\circ}$ lon-lat coordinate corresponds to the substellar point.}
      \label{fig:condense}%
    \end{figure}
Our simulations do not include CO$_2$ condensation. In the runs at 1 and 10~bars, the pressure found in a steady state never exceeds the vapor pressure of CO$_2$. In the 0.1~bar case, however, significant supersaturation is found at the surface and in the first 200~m above in some regions of the night side.  Figure~\ref{fig:condense} shows the areas and altitudes where the pressure exceeds the CO$_2$ vapor pressure. Although the formation of surface CO$_2$ ice and CO$_2$-ice clouds in the regions shown in Fig.~\ref{fig:condense} would have a negligible effect on the disk-integrated thermal emission, these regions could eventually trap the whole atmosphere into a surface deposit of CO$_2$ ice. A steady state would imply a release of CO$_2$ balancing the CO$_2$ loss into this cold trap, which may not be realistic in terms of timescales and which represents an \textit{ad-hoc} situation. Atmospheres modeled without CO$_2$ condensation require a minimum CO$_2$ pressure of about 0.5 bars to remain fully unsaturated. The 0.1~bar case presented in this study is thus not self-consistent. Simulations for 0.5 bars produce phase curves very similar to those obtained with 1~bar. It is possible that a pure CO$_2$ atmosphere could only exist for pressures above 0.5~bars (for the stellar, orbital and planetary parameters considered here). This would mean that once the planet is synchronized, its atmosphere must have already been outgassed, otherwise it would not be able to build up because the released gas would end up as CO$_2$ ice on the night side. Analysis of this interesting issue is beyond the scope of this study. Nonetheless, we decided to keep the $0.1$~bar case, for two reasons: \\
\noindent - Our paper focuses on the information contained in the spectral phase curves that can be measured for nontransiting planets, and comparison of the four chosen cases (0, 0.1, 1, and 10 bars) illustrates the effects of increasing pressure on the spectral and spatial distribution of the thermal emission. \\
\noindent - When assuming a pure CO$_2$ atmosphere and neglecting CO$_2$ condensation, we find that the mean surface pressure must be higher than $\sim 0.5$~bar to avoid supersaturation in the dark hemisphere. However, this minimum pressure is expected to be weaker if condensation is actually included with all its associated effects, in particular the release of latent heat and the radiative effect of CO$_2$-clouds. As the CO$_2$-ice particles form on the night side and blanket the surface thermal emission, both these effects would tend to warm the region where condensation occurs. Therefore, 0.5~bar is probably an overestimate of the minimum stable CO$_2$ pressure. \\
This link between CO$_2$ condensation and the stability of the atmosphere against the night-side cold trap is an interesting subject for future study.\\
\subsection{Non synchronized planets} \label{subsec:nonsync}
We restricted our study to a tidally-locked planet on a circular orbit. Atmospheric thermal tides can result in an equilibrium rotation period that is different from the synchronized one \citep{2008A&A...488L..63C}. This is for instance at the origin of the current rotation state of Venus \citep{2001Natur.411..767C}. Therefore, assuming a synchronized planet may not be consistent with a dense atmosphere.  Correia et al. (2008) show that the ratio $\omega/n$, where $\omega$ is the rotation rate and $n$ the mean motion, scales as  $\left ( a M_{*} \right )^{2.5}$ for a given planet and atmosphere. This ratio is 1.92 for Venus. A Venus-like planet at 0.05~AU  from a 0.31~M$_{\sun}$ star would thus have a ratio of as small as $5 \times 10^{-5}$ (which means $\sim 440$~days between sunset and sunrise).  Our simulated planet is 11 times more massive than Venus, while its $10$~bar atmosphere is $\sim 30$ times less massive than that of Venus, which would make the ratio $\omega/n$ even smaller. The departure from synchronization is therefore expected to be far too small to affect our results.
\subsection{Photon-noise limited observations}
In this paper, the only source of noise we consider is the stellar photon shot-noise, which is clearly a preliminary first step towards a realistic assessment of the observability of phase curves. The astrophysical background and zodiacal light should also be included in a realistic noise estimation. For JWST, one should take available estimates of the known sources of noise (thermal, read-out, dark current) into account for the instruments suitable for spectro-photometric observations, in particular MIRI and NIRSpec. Including these sources of noise should typically yield a factor 2 increase in the noise level (Belu et al., 2010). The stability of the instruments over one or more orbital periods is a key point that should be addressed in depth considering the required photometric precision ($10^{-4}$ to $10^{-5}$). EChO is a 1.5~m telescope proposed to the program Cosmic Vision 2 of ESA and dedicated to the characterization of exoplanet by combined star+planet light spectro-photometry. The aim for EChO is to achieve photon-noise limited observations and high detector stability over long periods. Our signal-to-noise basic estimates are therefore more relevant to the case of EChO. \\
\subsection{Stellar variability}
Stellar variability is key issue that cannot be solved directly by improving the instrumentation. The intrinsic stellar photometric variations and their wavelength-dependence, at a level of $\sim 10^{-5}$, represent an obvious impediment to the observation of nontransiting terrestrial planet phase curves. Basri et al. \citeyearpar{2010ApJ...713L.155B} studied the photometric variability of $\sim$150,000 Kepler stars over 33.5 days with a cadence of 30 min. This work shows  that fewer than about 20\% of main-sequence stars exhibit a variability lower than $\sim 10^{-3}$, and this fraction tends to decrease for low-mass dwarfs. What can contaminate the planetary phase curve, though, is the stellar component that varies periodically with the known period of the planetary modulation. This means that a minimum of two orbital periods will have to be observed in most cases in order to filter out anything that is aperiodic or periodic at another period. In-depth characterization of the stellar variability, including its origin, periodicity, and wavelength-dependence, is certainly required prior to any attempt to measure the phase curves of terrestrial planets. On an optimistic note, the optical phase variations of Kepler~10~b with a peak amplitude of only a few ppm has been measured, showing that an accuracy of 10 ppm must be achievable, at least in some cases. Our expertise is primarily in planetary physics, and we encourage experts of stellar variability to use their models and observations to assess the actual observability of the planet signature.
\subsection{Orbital distance and stellar types}
An 8-day period planet around an M3 dwarf (the case modeled in the present study) is not the most favorable case for attempting a phase curve observation of a terrestrial exoplanet. Hotter planets, with a shorter orbital period, are frequent and can provide better targets, depending on the stellar type and the wavelength. Known terrestrial exoplanets can be as hot as a few thousand K, such as Corot-7b \citep{2009A&A...506..287L} and Kepler-10b \citep{Batalha2011}, and Table~\ref{knownPlanets} shows that planets with terrestrial masses that are hotter than our model are known in the immediate neighborhood of the  Sun. 
\begin{figure}[hbt]
 \centering
\includegraphics[width=\linewidth]{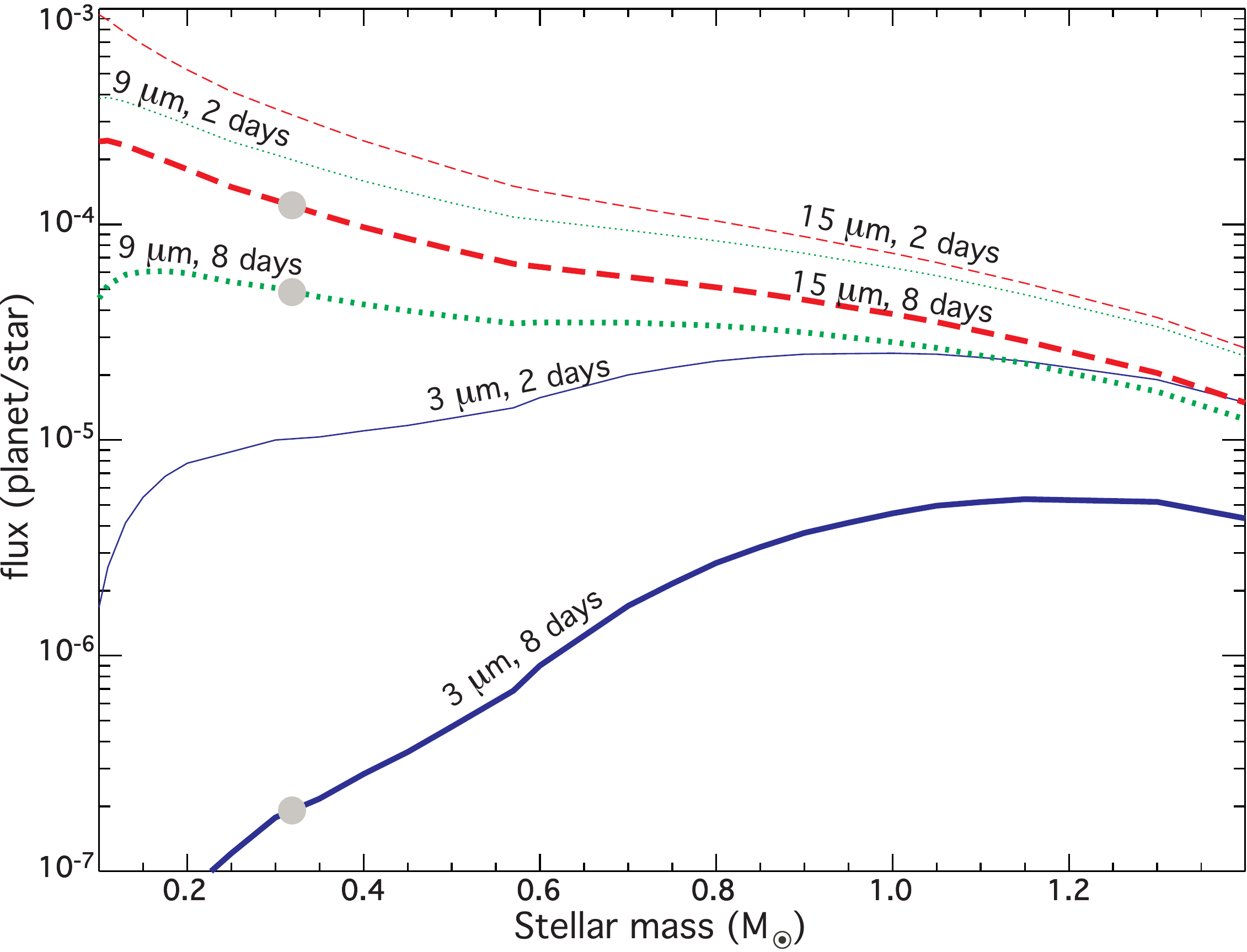}
 \caption{Amplitude of the phase curve for an airless planet as a function of stellar mass, orbital period and wavelength. The albedo of the planet is 0.2 and the inclination is 60$^{\circ}$. Grey circles indicate the case modeled in this paper. Stellar radii and T$_{eff}$ are from \citet{1998baraffe}. }
      \label{fig:contrast}%
    \end{figure}
Figure~\ref{fig:contrast} shows how the peak amplitude of the thermal phase curve (given in planet-to-star flux ratio) varies with the stellar type for two fixed orbital periods (2 and 8 days) and three wavelengths (3, 9, and 15~$\mu$m). One can see that phase curve at wavelengths around 3~$\mu$m can only be measured for very short-period objects around Sun-like stars. The search for the 2.7 and 4.3~$\mu$ CO$_2$ signatures may be restricted to such systems. We chose to model a planet below 600~K because one can argue that dense atmospheres may be rare around hotter planets due to atmospheric erosion. However, the existence and nature of dense atmospheres on terrestrial exoplanets, and their frequency as a function of orbital distance, stellar type, planet mass, must be eventually addressed by observation. It is therefore important to also target very hot planets, keeping in mind that the multiband phase curves are easier to measure in the absence of an atmosphere due to the high amplitude of the phase variations. For airless planets, multiband phase curves can be used to constrain the orbital inclination, the planetary radius and the surface albedo \citep{2011maurin}.
\subsection{Realistic atmosphere composition and their influence on phase curves}
As said previously, the atmosphere prototype that we used in this study is most certainly too simple compared to any of the realistic compositions that we can expect. Within the limited sample of known cases, the Venus atmosphere may represent an analog of some hot terrestrial exoplanets atmospheres. On Venus, the presence of clouds and the extremely dense atmosphere would make detecting the phase curve difficult, as all infrared wavelengths probe high altitude levels with little day-night variation. But even in such a case, if the signal-to-noise ratio were good enough, it would be possible to infer the presence of a dense atmosphere from not detecting the phase variation, as attempted by \citet{2009ApJ...703.1884S}. The ability to distinguish planets with dense atmospheres from airless bodies is essential for addressing the questions of atmosphere formation and survival at short orbital distances.\\
In future simulations, we will explore in detail the phase curve signatures of more complex atmospheres, containing a mixture of absorbers, condensable species, and aerosols.
\subsection{Null detection and albedo}
For nontransiting planets, the absence of photometric modulation can come from the presence of a dense atmosphere, to a very low inclination (see next section), or to a Bond albedo close to 1 (highly reflective planet). Such a high surface albedo is known only for ices, which can be ruled out for the starlit hemisphere of hot planets. Clouds can possibly produce a high Bond albedo (close to 0.7 for Venus), but cloud cover is also a signature of an atmosphere.  \\
We should, however, remain careful about the properties of planetary surfaces of unknown composition. For instance, the phase variation of the transiting planet Kepler~10b has been observed in the 0.4-0.9~$\mu$m band of Kepler \citep{Batalha2011}. This phase variation cannot be attributed to the thermal emission of the planet, which remains too low even for a null albedo. To fit the phase variation of this exoplanet when assuming no atmosphere and Lambertian scattering requires an albedo over 0.5 (possibly close to unity). One explanation could be that the Lambertian approximation cannot be applied at optical wavelength and that the actual scattering phase function peaks towards the star, hence towards the observer at the maximum of the phase curve. For extremely hot planets like Corot~7b and Kepler~10b, the existence of a surface magma ocean underneath a tenuous transient atmosphere is possible in the substellar area \citep{Leger2011}. The optical and reflective properties of such a surface are unknown. \\
In summary, a Bond albedo close to unity can be responsible for an undetected thermal emission but no known solid, non-icy planetary surface has this property. %
\subsection{Null detection and inclination}
With an inclination close to 0$^{\circ}$, the observer always sees half of the day- and night-sides, and the day-night brightness temperature contrast does not produce phase variations. In this configuration, periodic variations in the thermal emission can only be due to seasonal changes, because of either an obliquity or an eccentric orbit. Tidally evolved planets on circular orbits would not exhibit seasonal variations. \\
\begin{figure}[htb]
 \centering
\includegraphics[width=\linewidth]{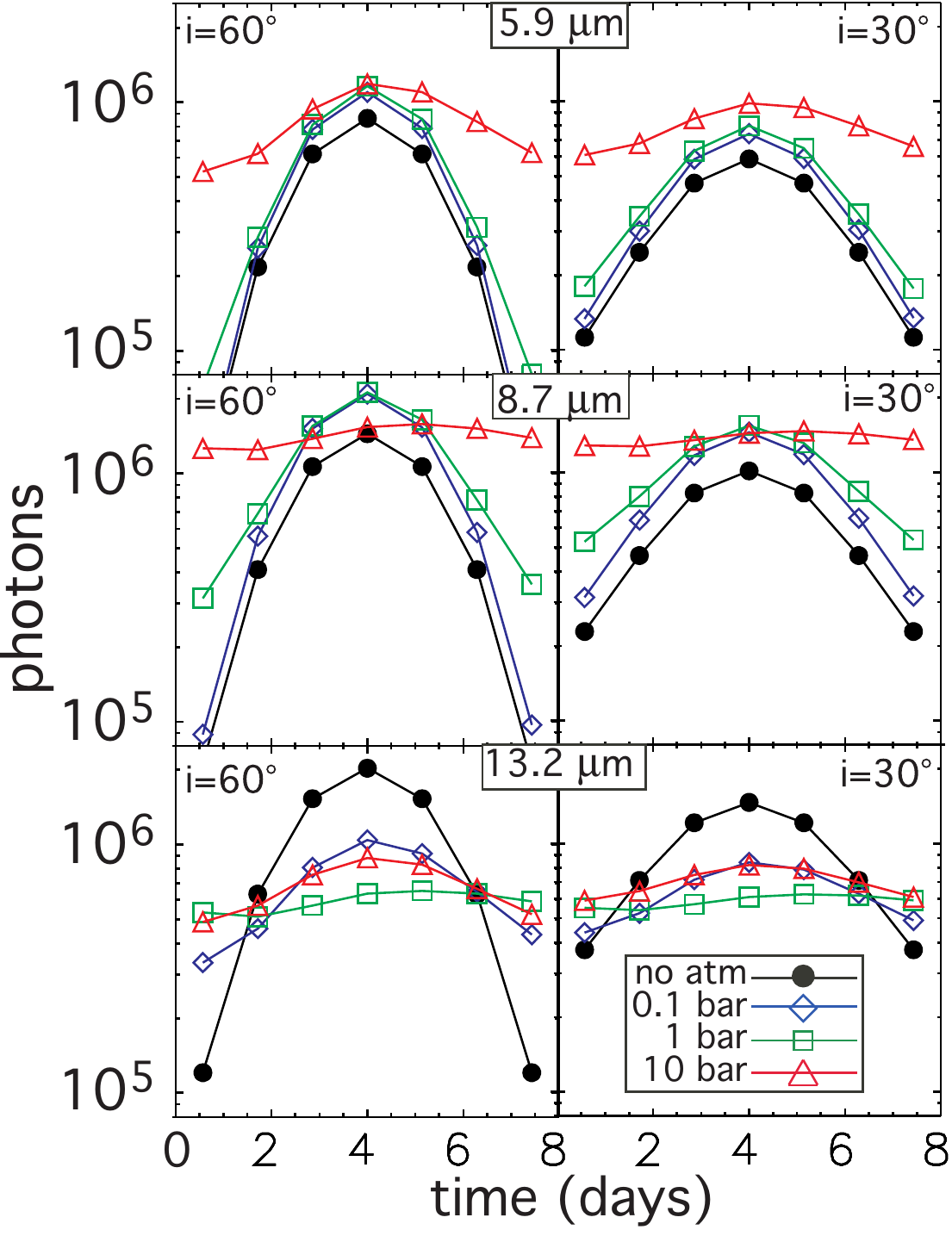}
 \caption{Thermal phase curves at 3 wavelengths for a 60$^{\circ}$ (left) and a 30$^{\circ}$ (right) inclination. Photon fluxes are calculated for a 1.5~m telescope. Graphs on the left are the same as in Fig~\ref{fig:phaseEChO}. }
 \label{inclination}
 \end{figure}
The probability of a very low inclination is, however, low (only 6\% and 1.5\% of randomly oriented systems have an inclination lower than 20$^{\circ}$ and 10$^{\circ}$, respectively) and the sensitivity to inclination is weak for inclinations over 20$^{\circ}$. This weak sensitivity to the inclination is due to the small variation of $cos(\theta)$ (where $\theta$ is  the zenith angle) over a wide angular region around the substellar point. As a result, a large fraction of this hot substellar region contributes to the emission at the maximum phase even for inclinations down to about 20$^{\circ}$. Figure~\ref{inclination} compares phase curves obtained at 60$^{\circ}$ and 30$^{\circ}$ inclinations. In addition, the range of possible inclinations can often be constrained. When the measured projected rotation velocity of the star $Vsin(i)$ is measured, a minimum value for $i$ can be deduced from the possible range of values for $V$ for the type and age of the star, and $V$ can also be measured for active stars by the photometric variability due to the stellar spots. This was done for the planet-hosting star GJ876 \citep{2010A&A...511A..21C}. This reasoning assumes that the $i$ in $Vsin(i)$ is close to the orbital inclination. This can be very wrong for hot Jupiters, but it might be more reasonable for other systems. In multiple systems the orbital inclinations can be constrained directly when fitting dynamical orbits to the radial velocity data, as done for planets $b$ and $c$, also around GJ876 \citep{2010A&A...511A..21C}. In the GJ581 system, where the hot terrestrial planet $e$ would be a fine target for phase curve observation,  inclinations lower than 60$^{\circ}$ would not be consistent with a stable system, due to the resulting high planetary masses and associated planet-planet interactions (assuming that mutual inclinations between the planets remains small). Astrometric measurements could also be used to determine the inclination, or at least constrain it, although the short-period planets we consider here are not easy targets for astrometry.\\
The number of configurations in which a low inclination prevents us from inferring any atmospheric properties should thus remain low, and these cases are possible to identify.
\subsection{Validity of the Lambertian approximation} \label{subsec:lambert}
In each latitude-longitude cell, the flux $F_\lambda$ at the top of the atmosphere for each band is an output of the GCM. To compute the flux received by an observer, we need the angular distribution of specific intensities. We assume an isotropic (Lambertian) distribution of intensities ($I_\lambda=F_\lambda /\pi$). This approximation neglects limb darkening (or limb brightening, which is possible when a temperature inversion exists).  To quantify this approximation, we used a detailed 1D line-by-line radiative transfer code.  We considered an uniform atmosphere (same P(z) and T(z) everywhere) and computed the specific intensities at the top of the atmosphere with a resolution of $5^{\circ}$. We used these specific intensities to directly compute the fluxes received by a remote observer, and we also integrated them to calculate the top-of-the atmosphere fluxes. From these top-of-the atmosphere fluxes, we calculated the approximated disk-integrated fluxes received by the distant observer, assuming an isotropic distribution of intensities. Comparison of the two disk-integrated fluxes showed that the error due to limb-darkening is below 5\% in all our bands. \\ It is, howeve,r likely that the error could be more significant for a non uniform atmosphere. Indeed, the hot substellar region is responsible for the largest phase curve modulations. At some wavelengths, this region should be fainter than it is in our modeling when it is near the edge of the apparent planetary disk. At low inclinations ($<30^{\circ}$), this could lower the amplitude of the variations. At higher inclinations, the amplitude of the modulation should not be significantly affected, but the variation should be less sharp at wavelengths where the visibility of the hot spot decreases when approaching the edges of the planetary disk. Doing the full line-by-line spectrum synthesis in 3D would be too time-consuming but, for future studies, we are developing a 3D band model at the resolution of the GCM.
\subsection{Transiting planets}
Phase curves can also of course be measured for systems in a transit configuration. In this case, the absolute flux and not just the variations can be obtained, as the flux of the star alone is inferred from the secondary eclipse. In addition, the day-side emission spectrum of the planet can be measured by comparing the star + planet emission just after or before the secondary eclipse with the emission during the secondary eclipse. However, as in the nontransiting configuration, the phase curve provides precious information on the atmospheric circulation, with higher accuracy thanks to the ability to measure the absolute planetary flux. In addition, the variation spectrum is not limited by the transit duration. For instance, the transit duration for the system studied in this paper is 1h 30min, while the phase curve can be sampled with seven exposures of $\sim$ 28h to cover the 8-day orbit. The phase curve and secondary eclipse obtained on Kepler~10b  illustrate this point nicely. Although in this case it is the scattered light that has been observed, Fig.~13 of Batalha et al. \citeyearpar{Batalha2011} shows that the amplitude of the phase curve is better measured than the depth of the secondary eclipse.
\section{Conclusions}
By using 3D climate simulations of pure CO$_2$ atmospheres on a tidally-locked, moderately hot terrestrial planet, we have studied the influence of an atmosphere on the multiwavelength thermal phase curve. We described the effect of increasing pressures on the spectral and spatial distribution of the thermal emission and on the observed phase curves.\\

We have shown that the \textbf{variation spectrum} (the peak amplitude of the observed modulation due to the planet as a function of wavelength), which can be measured for a nontransiting planet, is shaped by molecular bands, just like the emission spectrum. This is because the day-night temperature contrast changes with altitude, and different wavelengths probe different altitudes. Signatures of atmospheric molecules can thus be extracted from spectro-photometric observations of nontransiting planets. \\

The existence of an atmospheric window probing the near-surface layer is, however, necessary for obtaining enough phase curve variations. A high atmospheric column density, a mixture of absorbers covering the whole wavelength range of the observations, or clouds, could prevent such windows and restrict the characterization to the sole inference of a dense atmosphere.  \\

The 4.3~$\mu$m band of CO$_2$ appears in the variation spectrum at all pressures and is thus a promising way to detect CO$_2$. However,  measuring the phase curve at this wavelength would require a hotter planet than the one studied here, in order to reach observable planet/star contrasts and SNR. \\

On an 8-day synchronous planet around an M3 dwarf, a pure CO$_2$ atmosphere is stable against a condensation collapse but only for mean surface pressures above a certain threshold, estimated to be in the 0.1-0.5~bar range. In a future study, we intend to refine the actual value of this minimum pressure and its dependence upon parameters such as surface albedo, stellar type, orbital period, and planetary radius. \\

A photon-noise limited instrument of only 1.5~m with an extremely stable detector able to perform photometry at a level of $10^{-5}$ over several days simultaneously in several spectral bands could in theory detect modulation by a nontransiting planet with a precision allowing the atmosphere to be characterized. The level of characterization that can be achieved depends on the nature and density of the atmosphere. \\

The wavelength-dependent phase curves computed in this study can be used to assess the observability of nontransiting terrestrial exoplanets by taking intrinsic stellar variability into account. The stellar variability and photometric stability of the instrument are certainly the main challenges when measuring the phase curves of nontransiting and spatially-unresolved terrestrial exoplanets.

\begin{acknowledgements}
      FS acknowledges support from the European Research Council
(Starting Grant 209622: E$_3$ARTHs).
\end{acknowledgements}

\bibliographystyle{aa}




%

%

\end{document}